# Emerging orthorhombic two-dimensional van der Waals magnets

Xiaoyu Guo[1,+], Wenhao Liu[2,+], Bing Lv[2,*], Liuyan Zhao[1,*]

[1] Department of Physics, University of Michigan, Ann Arbor, MI 48019, USA
[2] Department of Physics, the University of Texas at Dallas, Richardson, TX 75080, USA
[*] Corresponding authors: blv@utdallas.edu (B. Lv); lyzhao@umich.edu (L. Zhao)
[+] These authors contribute equally.

## Abstract

Two-dimensional (2D) magnetism realized in van der Waals (vdW) materials has expanded to include a great variety of magnetic phases, over a short decade since its first discovery in 2016-2017. However, most of the investigated vdW magnets so far have highly symmetric crystal fields and isotropic in-plane lattice structures, making their 2D magnetism robust but often classical and expected. In this Perspective, we highlight a family of vdW magnets that have distorted crystal fields with low symmetries and orthorhombic crystal lattices with strong anisotropy, and furthermore, show their prospects for realizing a much richer landscape of magnetic phases and phase transitions. First, we introduce the classification of 2D vdW magnets based on their crystal fields and make a contrast between the orthorhombic vdW magnets of focus here and other popular honeycomb/hexagonal/kagome vdW magnets. We then discuss the material candidates, the emergent properties, and the designed controls for these orthorhombic 2D vdW magnets, using the representative CrSBr as an example and extending beyond it. Finally, we discuss a few existing challenges in orthorhombic 2D vdW magnets, overcome which would pave the way towards magneto-optical, opto-magnetic, and spintronic devices based on orthorhombic 2D vdW magnetism.

## Main Text

For a two-dimensional (2D) system, orthorhombicity describes the differentiation between the two orthogonal in-plane directions. From a symmetry perspective, orthorhombic crystal system is one of the seven crystal systems [1], containing three structural point groups and nine magnetic point groups [2]. A key manifestation of structural orthorhombicity in 2D and quasi-2D materials is to introduce anisotropy in other physical properties spanning from electronic, magnetic, optical, *etc.*, to magneto-electric, opto-magnetic, opto-electronic, *etc.*, properties. One can see such a manifestation from two levels, *i.e.*, the interaction and the response. On the microscopic interaction level, it refers to the different electronic hopping, magnetic exchange, or light-matter interactions, *etc.*, between the two in-plane crystalline directions [3–6]. In the extreme case, it is possible to have strong interaction along one direction and fully quenched interaction in the orthogonal direction, making a structurally 2D system effectively one-dimensional (1D) in certain physical properties [7–11]. On the macroscopic response level, it leads to the contrasting behaviors between the two in-plane crystalline axes even under equivalent external stimuli along these two axes. As a result, it often leads to highly directional responses that are desired in various device applications [6,12–14].

2D antiferromagnetism (AFM) and ferromagnetism (FM) in van der Waals (vdW) materials were first discovered in monolayer $FePS_3$ in 2016 [15,16] and monolayer $CrI_3$ [17] and few-layer $CrGeTe_3$ [18] in 2017, which marks the opening of the blooming field of 2D magnetism [19–21]. They have served as a platform to investigate phase transitions in the 2D limit and explore potential new phases of matter [22], offered the magnetic ingredient to construct novel heterostructures and exploit its proximity effect [23,24], and further provided ample opportunities to implement the concept of 2D magnet-based micro-spintronic devices [25]. All of these have motivated the rapid expansion of the 2D vdW magnet pool, and as of now, it includes 3*d* and 4*d* transition metal halides [26–36], 3*d* transition metal chalcogenides [37–

44], ternary 3d transition metal compounds [15,16,18,45–62], $Fe_nGeTe_2$ ($3 \leq n \leq 5$) [63,64], $MnBi_{2n}Te_{3n+1}$ ($n = 1,2,3,4$) [65–70], and even 5d- and f-electron based compounds [71–75].

| crystal field symmetry | energy levels | materials | building block | atomic lattice |
|---|---|---|---|---|
| $C_{2v}$ | 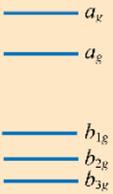 | MXY $\begin{pmatrix} M = Ti, V, Cr, Fe \\ X = O, S, Se, Te \\ Y = Cl, Br, I \end{pmatrix}$ | 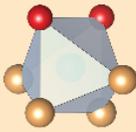 | Orthorhombic 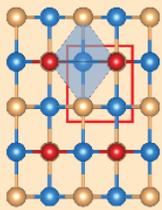 |
| $D_{3h}$ | 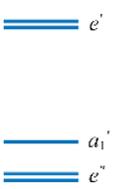 | 2H-phase $MX_2$ $\begin{pmatrix} M = V, Cr, Mn, Co \\ X = S, Se, Te \end{pmatrix}$ | 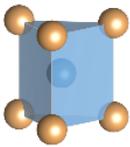 | Honeycomb 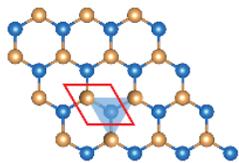 |
| $C_{3v}$ | 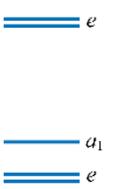 | $Nb_3X_8$ (X = Cl, Br, I) | 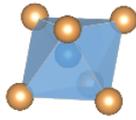 | Kagome 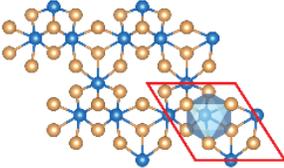 |
| $O_h$ | 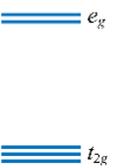 | $MX_3$ $\begin{pmatrix} M = V, Cr, Mn^*, Fe^*, Ni^* \\ X = Cl, Br, I \end{pmatrix}$<br>$MX_2$ $\begin{pmatrix} M = V, Cr, Mn, Fe, Co, Ni \\ X = Cl, Br, I \end{pmatrix}$<br>1T-phase $MX_2$ $\begin{pmatrix} M = V, Cr, Mn, Co \\ X = S, Se, Te \end{pmatrix}$<br>$CrXTe_3$ (X = Si, Ge, Sn)<br>$MPX_3$ $\begin{pmatrix} M = V, Cr, Mn, Fe, Co \\ X = S, Se, Te \end{pmatrix}$<br>α-$RuX_3$ (X = Cl, Br, I)<br>Fe-Ge-Te compounds<br>Mn-Bi-Te compounds | 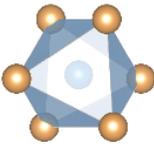 | Honeycomb 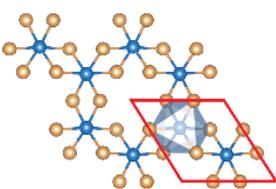<br>Hexagonal 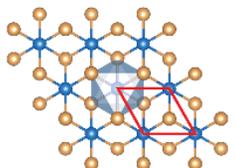 |

**Figure 1 | Classification of 2D magnets based on their crystal fields and crystalline lattices.** From the left to the right, each column represents the crystal field symmetry, the d-orbital degeneracy lifting under the crystal field, the discovered 2D magnetic materials, the atomic building block, and the lattice structure with the corresponding crystal system labeled. The red rectangle and parallelograms mark the unit cells.

A quick examination of the 2D vdW magnet pool reveals that most of their monolayer crystal lattices belong to the trigonal and the hexagonal crystal systems with in-plane isotropy (see **Figure 1**). In contrast to this majority, one family of 2D vdW magnets, MXY, stands out due to their orthorhombic crystalline lattices that allow for in-plane anisotropy, where M = transition metals, X = chalcogens, and Y = halogens. The well-known member of this family is CrSBr [76,77], which hosts a quasi-1D conduction band [77–79], a robust layered AFM order [61,80], a rich magnetic phase diagram [61,81–85], exceptionally strong light-matter coupling [86–91], giant exciton binding energy [80,92–96], and linearly polarized photoluminescence [79,89,92], as well as air stability and high Néel temperature [61,78,80–82,84,85,97–100]. A peek into these distinctive properties of CrSBr, most of which originate from the structural orthorhombicity, which convinces us and invites us to explore MXY, the family of orthorhombic vdW magnets.

## Classification of 2D vdW magnets based on crystal fields

The crystal lattice of a 2D vdW magnet is closely tied with its building block, which is the magnetic transition metal cation situating inside a cage made of ligand anions. Locally at the transition metal site, the *d* orbitals will experience the electrostatic potential created by the ligand cage, and as a result, its 5-fold degeneracy will be lifted in different ways depending on the symmetry of the cage. This crystal field effect is crucial in determining the spin state of the transition metal cation, the single ion and the exchange magnetic anisotropies for the 2D vdW magnet [101–103]. The crystal field is classified upon the point symmetry group at the transition metal site, and the common ones include the octahedral $O_h$, distorted octahedral $D_{2h}$ or $C_{2v}$, trigonal prismatic $D_{3h}$, triangular prismatic $C_{3v}$, tetragonal $T_d$ crystal fields, *etc*.

Nearly all vdW magnets thus far concern transition metals with *d* orbitals. **Figure 1** summarizes the crystal fields and the corresponding *d* orbital degeneracy lifting, with the vdW magnet candidates listed based on their crystal fields and crystal systems. From this summary, it is evident that most of the known vdW magnets have the $O_h$ crystal field, the one with the highest symmetry and *d* orbitals splitting into the doubly degenerate $e_g$ and triply degenerate $t_{2g}$ manifolds. Following $O_h$, they are the $D_{3h}$ and $C_{3v}$ crystal fields with the *e* doublets protected by the 3-fold rotational symmetry. Finally, it is the $D_{2h}$ and $C_{2v}$ crystal field with the lowest symmetry and the fully lifted degeneracy for the *d* orbitals. It is also illuminating to notice that the vdW magnets with the 3-fold rotational symmetric crystal fields (*i.e.*, $O_h$, $D_{3h}$, $C_{3v}$) also fall into the isotropic crystal systems (*i.e.*, trigonal, hexagonal). Of interest to this Perspective is the family of MXY, which has the least symmetric $C_{2v}$ crystal field and the anisotropic orthorhombic crystal system.

## Distorted crystal field and orthorhombic lattice for enriched physical properties

The crystal structure of MXY is the same as that of CrSBr, each layer of which consists of transition metal monochalcogen MX double layer sandwiched between layers of halogen Y. The structural primitive cell contains two edge-sharing distorted octahedra, with one $M^{3+}$ cation in the center and four $X^{2-}$ and two $Y^-$ anions at the vortices for each octahedron, which then arranges into an in-plane (*ab*-plane) orthorhombic network and further stacks vertically along the out-of-plane (*c*-axis) direction (**Figure 2**, center bottom) [80,104].

Within the distorted MXY octahedron, the coordination of two types of anions around the metal cation breaks the spatial inversion symmetry and leads to the low-symmetry $C_{2v}$ crystal field. Comparing to the high-symmetry crystal fields, *e.g.*, $O_h$, $D_{3h}$, $C_{3v}$, this $C_{2v}$ one has its own unique advantages. First, the $C_{2v}$ crystal field fully lifts the orbital degeneracy of the five 3*d* orbitals, and thus is beneficial for stabilizing the spin state when tuning across $3d^1$ to $3d^5$ via changing $M^{3+}$ from $Ti^{3+}$ to $Fe^{3+}$ [105–114]. This enables the possibility of exploring the magnetic phase diagram of varying spin states under the isostructural condition. Second, the strong distortion of the $X_4Y_2$ octahedral cage surrounding $M^{3+}$ makes the electronic hopping and magnetic exchange coupling especially sensitive to the bond angles and length, and therefore empirically correlates the lattice, charge, and spin degrees of freedom (**Figure 2**, center middle).

This brings the opportunities of efficiently controlling the electronic and magnetic phases via applying static strain to the crystal lattice [115–123].

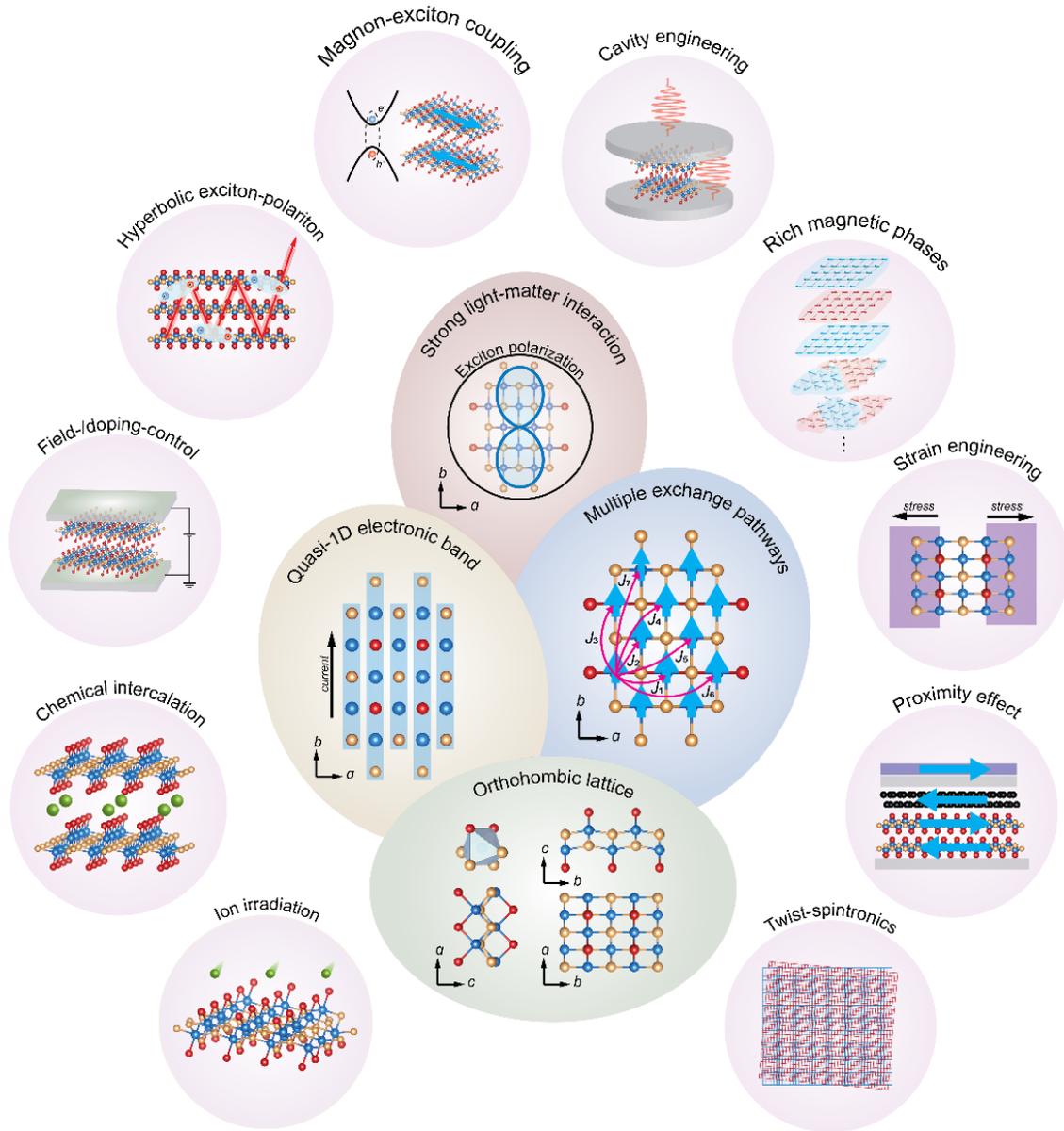

**Figure 2 | Overview of key properties of 2D orthorhombic magnets and their associate opportunities.** In the center, four unique physical properties of 2D orthorhombic magnets are shown, regarding the lattice, charge, and spin degrees of freedom and the coupling to light. In the outer shell, a series of research directions based on the four properties are highlighted.

The in-plane orthorhombic lattice of MXY can be seen from a different perspective – that is chains of metal cation $M^{3+}$ with $X^{2-}$/ $Y^{-}$ ligands running along the $a$-axis to form ribbons of MXY and then these MXY ribbons connecting perpendicular to the $a$-axis to form a layer in the $ab$-plane. The M-M distance is ~ 3Å along the chain direction and ~ 5Å between ribbons, showing a clear, strong in-plane anisotropy for this orthorhombic crystal lattice. As compared to the in-plane isotropic crystal lattices, *e.g.*, trigonal and hexagonal, this orthorhombic one brings distinct flavors. First, this in-plane lattice anisotropy can lead to highly anisotropic electronic band dispersion (**Figure 2**, center middle), or even quasi-1D electronic band such as the case of CrSBr whose conduction band is only dispersive along $\Gamma - Y$ but dispersionless along $\Gamma - X$ [92–96,117,118,124]. Such a flatband further leads to exceptionally strong light-matter interactions in CrSBr in free space [86,92] and inside photonic cavities [87–91] (**Figure 2**, center top), which suggests that light and/or photonic cavities can be efficient knobs to tune the physical properties of MXY. Second, the structural and electronic anisotropies can further lead to different types and strengths of direct exchange and superexchange interactions along different crystal axes (**Figure 2**, center middle), and in the extreme limit, could lead to the quasi-1D spin chain state [107,125]. This offers an unprecedented opportunity of studying a full spectrum of dimensionality crossover from 3D to quasi-2D to 2D to quasi-1D. A survey of exciting research opportunities stemming from orthorhombic lattice, quasi-1D electronic band, complex magnetic exchange, and strong coupling to light are highlighted in the outer ring of **Figure 2**.

## Material candidates of MXY vdW magnets

In principle, the family of MXY vdW magnets include M = Ti, V, Cr, Mn, and Fe; X = O, S, Se, and Te; and Y = Cl, Br, and I. In practice, the oxyhalides, X = O, contribute the largest group of compounds that have been successfully synthesized, whereas the X = S group has CrSBr, and the X = Se, Te groups have no known compounds in the single crystal form thus far (**Figure 3**).

For the X = O group, the magnetic transition metal cation $M^{3+}$ varies from $Ti^{3+}$ to $V^{3+}$, $Cr^{3+}$, and $Fe^{3+}$, and hence allows for the 3$d$ electron configurations from $d^1$ to $d^2$, $d^3$, and $d^5$. We note that there has been no report on MnOY with the $Mn^{3+}$ cation and the $d^4$ configuration, possibly due to the disproportionation reaction from $Mn^{3+}$ to more stable $Mn^{2+}$ and $Mn^{4+}$ cations. The research on transition metal oxyhalides started with the synthesis of magnetic transition metal oxycholorides (MOCl with M = Ti, V, Cr, and Fe) by Schafer *et al* back in the late 1950s to the early 1960s [126–129], and then was further pursued in detailed structural analysis in the late 1970s [130–132], nearly two decades after the initial synthesis. It was until the recent 2000s when comprehensive magnetic and transport measurements were carried out for these oxycholorides polycrystals and single crystals [105,108–110,114,133–135], and even more recently in the 2020s when their magnetic properties in the 2D limit are explored [136,137]. In sharp contrast to oxycholorides (MOCl), oxybromides (MOBr) and oxyiodides (MOI) were much less studied, with only a couple of structural and magnetic characterizations for TiOBr in the early 2000s [125,138]. It is an interesting topic to investigate the magnetism evolution upon the variation of transition metal cations and the halogen anions in MOY.

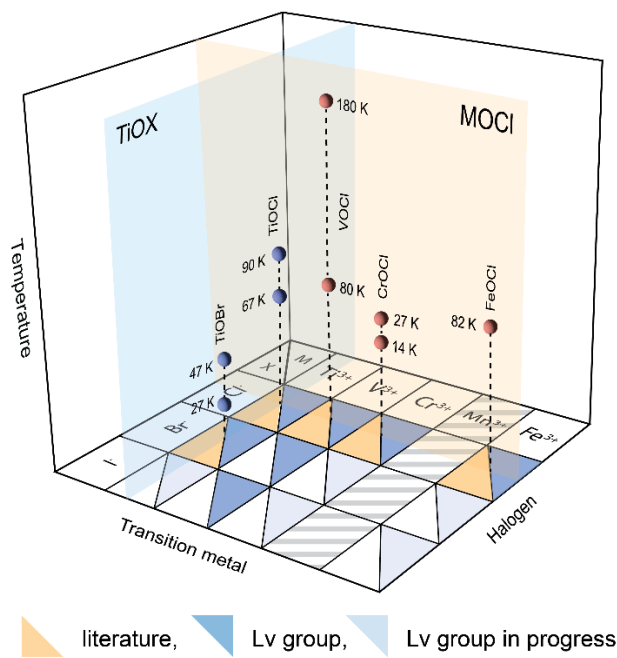

**Figure 3 | Summary of magnetic phase transitions in MOY.** A table of possible MOY magnetic materials is shown, with the literature reported ones marked in orange, the Lv group synthesized ones marked in blue, and the Lv group in progress ones marked in lighter blue. Magnetic phase transitions as a function of temperature are shown for the families of MOCl (M = Ti, V, Cr, Fe) and TiOY (Y = Cl, Br).

Let's first take oxycholorides as an example to see the effect of transition metal cations (**Figure 3**). TiOCl first undergoes a magnetic phase transition across $T_{c1}$ = 90 K into an incommensurate spin state and then a first-order phase transition at $T_{c2}$ = 67 K into a spin-Peierls state, both of which manifests as anomalies in temperature dependent magnetic susceptibility [105–107]. VOCl was reported to have a first magnetic phase transition at $T_{c1}$ = 180 K whose nature remains elusive and unexplored [109] and then a second magnetic transition at $T_{c2}$ = 80 K into an AFM state with a concurrent monoclinic structural distortion [108,109]. CrOCl also experiences two magnetic phase transitions, a higher one at $T_{c1}$ = 27 K that is possibly due to an incommensurate magnetic order and a lower one at $T_{c2}$ = 14 K that is for an AFM order accompanied with a monoclinic structural transition [110]. FeOCl transitions into an AFM state at $T_c$ = 82 K together with a monoclinic structural distortion [98]. The spin-Peierls state in TiOCl and the monoclinic structural transition together with the AFM order in VOCl, CrOCl and FeOCl suggest the strong magnetoelastic coupling present in this group of oxycholorides [110,114,135], echoing that strain shall be a promising control knob for the magnetism in MOCl. Furthermore, the two magnetic phase transitions in TiOCl, VOCl, and CrOCl, together with the metamagnetic phase transitions found in CrOCl under the magnetic field [113], indicate the strong magnetic frustrations in MOCl, providing a new platform to investigate the interplay amongst geometric frustration, critical fluctuations, and collective phenomena in the 2D limit.

Let's then take a look at the effect of halogen ligand anions on the magnetism, where TiOY (Y = Cl, Br, I) is used as an example (**Figure 3**). TiOBr host a similar process of two magnetic phase transitions as TiOCl, but at lower critical temperatures of $T_{c1}$ = 47 K into an incommensurate magnetic order and $T_{c2}$ = 27 K into a spin-Peierls state [139]. TiOI has been hardly synthesized thus far. The similarity in magnetic phase transitions between TiOCl and TiOBr suggests the robustness of the magnetism to the halogen ligand anions.

For the X = S group, the only known compound is CrSBr [61,80,104], the material that has motivated this perspective. CrSBr undergoes a multi-stage magnetic phase transition process to enter the layered AFM order [61,82,100], supports the long-sought extraordinary phase transition [83], hosts a quasi-1D electronic conduction band [77–79,93–96], anisotropic excitons and polaritons [86–92,140], and strong spin-exciton coupling [89–91,140–144] (see details in the section of Emergent Properties in MXY vdW Magnets), as well as responding sensitively to external controls including strain [115,117–123], pressure [145,146], doping [147,148], electric field [149], twist [150–152], intercalation [153,154], ion and electron irradiation and implantation [124,155,156], *etc*. (see details in the section of Designed Controls over MXY vdW Magnets).

For the X = Se, Te groups, there has been no successful synthesis thus far. **Figure 3** summarizes the synthesis of MXY vdW magnets that have been reported in literature and/or produced by the Lv group.

## Emergent properties in MXY vdW magnets

The distorted crystal field and the anisotropic orthorhombic lattice prepare MXY vdW magnets with a variety of exciting emergent properties. CrSBr is the most investigated member of the family thus far, and it has already shown distinct electronic, excitonic, polaronic, and magnetic properties.

***Quasi-1D electronic conduction band.*** CrSBr is a semiconductor with the Fermi level lying within the semiconducting gap. Angle resolved photoemission spectroscopy (ARPES) can only access the electronic conduction band of CrSBr when CrSBr is doped by electrons, for example, through the charge transfer between the CrSBr flake and a metallic Ag(111) or Au(111) substrate [95] or the surface doping by the deposition of K atoms [93]. For the K-doped CrSBr bulk, the electronic conduction band is nearly dispersionless along the $\Gamma - X$ direction, but clearly dispersive along the $\Gamma - Y$ direction, as shown in **Figure 4a**, which is direct evidence of the quasi-1D nature and the flat-band character of the conduction band. It is further worth noting that the electron doping in CrSBr does more than simply raising the Fermi

level – it causes a significant reduction of the electronic band gap in the K-doped CrSBr [93] and even drives a Lifshiftz transition for the electronic band of CrSBr flake on Ag (111) [95].

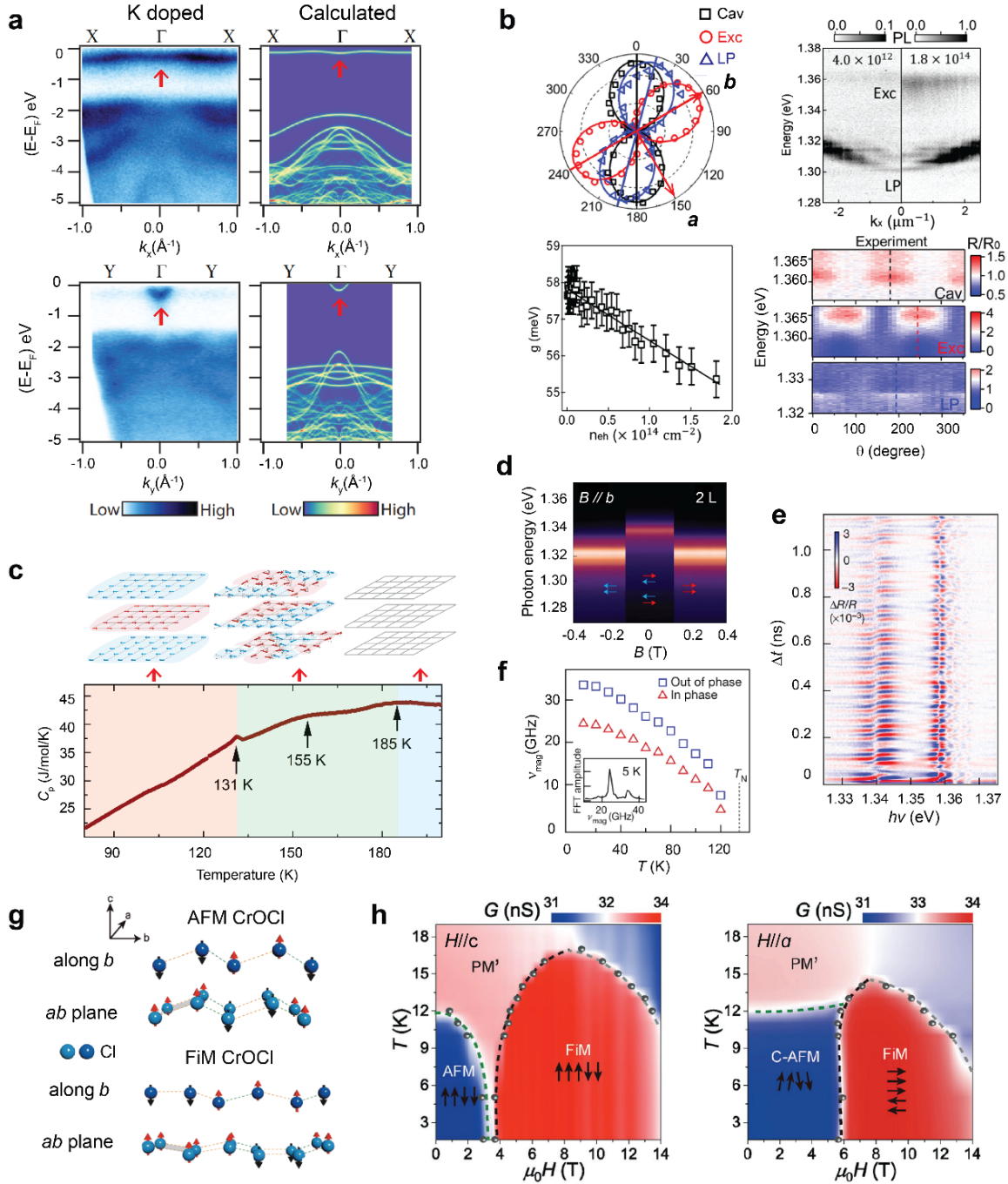

**Figure 4 | Emergent properties in MXY vdW magnets. a.** Quasi-1D electronic band of K-doped CrSBr that is measured by ARPES and compared with theoretical calculations [93]. **b.** Linear polarization of the exciton mission in CrSBr and strong light-matter coupling of CrSBr integrated into photonic cavities [89]. **c.** Multi-stage magnetic phase transitions in bulk CrSBr shown by the heat capacity measurements [82]. **d-f.** Strong charge-spin coupling shown in CrSBr through the exciton energy shift upon the layer AFM to FM phase transition (**d**) [92] and the coherent population of magnons (**e, f**) [141]. **g-h.** Complex magnetism and magnetic phase diagram of CrOCl [113]. The figures in (**a-h**) are adapted from Refs. [93,89,82,92,141,113].

***Anisotropic excitons and polaritons.*** The quasi-1D electronic conduction band readily contributes to the giant binding energy [80,92–96], the large oscillator strength [89], and the strong linear polarization of excitons in CrSBr [79,88,89,91,91,92,140]. Despite the variation in the electronic bandgap reported in literature using STM [79,80], ARPES [93–96], and optics [91,92], the photoluminescence frequency of CrSBr excitons is consistently within the range of 1.25–1.38 eV [79,80,91,92,140,141,144,147,157,158] and its polarization is highly linear along the *b*-axis direction [79,88,89,91,91,92,140] (**Figure 4b**). Integrating the CrSBr flakes into photonic cavities, exciton-polaritons emerge with a strong exciton-photon coupling strength that brings the system into the strong-coupling regime and with an exceptionally large Rabi splitting that is greater than any known systems to date [86–89] (**Figure 4b**). Furthermore, the polarization of the exciton-polariton inherits the linear polarization from the exciton, and at the same time, can be tuned by the photonic cavity polarization (**Figure 4b**) [89].

***Multi-stage magnetic phase transitions.*** The magnetic ground state of bulk and few-layer CrSBr is the layered AFM order where the spins align ferromagnetically along the *b*-axis within the layer and antiferromagnetically between the adjacent layers (**Figure 4c**) [61,80,100,104]. The Néel temperature for bulk CrSBr is $T_{N,bulk}$ = 132 K confirmed by the heat capacity and the magnetic susceptibility measurements [61,80,82,85,104]. Prior to $T_{N,bulk}$, three more characteristic temperature scales have been found for bulk CrSBr, $T_{N,surf}$ = 140 K as the critical temperature for the surface layered AFM order [83], $T^*$ = 155 K as a characteristic temperature for either a *c*-axis incoherent ferromagnetic order or a short-range layered AFM order [61,82,83,100], and $T^{**}$ = 180 K as a temperature scale for the development of spin-spin correlations [82] (**Figure 4c**). Such a rich multi-stage magnetic phase transition process distinguishes CrSBr from many honeycomb and hexagonal vdW magnets. Surprisingly, when thinning down CrSBr to the 2D limit, the magnetic critical temperature increases, $T_{N,2L}$ = 140 K for bilayer CrSBr and potentially $T_{C,1L}$ = 146 K for monolayer CrSBr [61,85]. This counter-intuitive trend of the magnetic onset temperature with respect to the layer number remains to be explored. At $T_F$ = 30-40 K, there are reports of a ferromagnetic phase transition, whose origin is still under debate [81,85,98,159].

***Strong charge-spin coupling.*** The spin and charge degrees of freedom are strongly coupled in CrSBr. The exciton energy has been shown to depend on the interlayer magnetic order, being 1.34 eV in the layered AFM phase and shifting down to 1.32 eV in the FM phase in bilayer CrSBr (**Figure 4d**) [92]. This sensitive dependence of exciton energy on the interlayer spin alignment further leads to an exciton frequency oscillation for about 4 meV when the spin waves are coherently excited (**Figure 4e**) [141]. Two spin wave modes were captured, one at 24 GHz for the in-phase spin precession between adjacent layers and the other at 34 GHz for the out-of-phase counterpart (**Figure 4f**) [141,160–162] . Given the large mismatch in energy between excitons and magnons, such a strong coupling between the two types of quasiparticles is exceptional.

***Magnetic frustration and metamagnetism.*** In addition to CrSBr, CrOCl receives increasing attention lately. Different from CrSBr, CrOCl realizes an AFM order within the layers that has a slightly complicated AFM spin arrangement of ↑↑↓↓ along the *b*-axis and a FM chain along the *a*-axis (**Figure 4g**) [112,163]. Such an intralayer AFM order is a direct consequence of the geometrical frustration of spins, which is why this AFM transition is accompanied by a monoclinic structural distortion [110,113]. Furthermore, this geometric frustration leads to metamagnetic phases (**Figure 4h**) [111,113] and even a liquid-gas type phase transition when an external magnetic field is applied, suggesting a rich magnetic landscape for CrOCl, or more broadly, for the oxycholoride group.

## Designed controls over MXY vdW magnets

The intimate coupling between the lattice, charge, and spin degrees of freedom in MXY suggests diverse possibilities for controlling the physical properties of MXY. For example, from the lattice perspective, one can use strain or pressure to modify the atomic structure or twist to introduce the superlattice; from the charge perspective, one can use electric field effect or chemical intercalation to tune the charge carrier density; from the spin perspective, one can use electric field to couple into the parity broken magnetic state or magnetic field to control the spins directly. In fact, these external controls have been successfully practiced in CrSBr.

***Uniaxial strain.*** The complex magnetic exchange pathways and the magnetic anisotropies of CrSBr are highly sensitive to the in-plane lattice constants. A uniaxial strain device provides an efficient way to modify the lattice constants and therefore the magnetic properties. By applying a uniaxial strain along the *a*-axis of a ~20 nm CrSBr flake, a magnetic phase transition from the layered AFM to FM happens at a critical strain of 1.5% (**Figure 5a**), showing the strain-induced modification of interlayer exchange coupling. Moreover, this tensile strain also causes a significant reduction in the anisotropy energy and tunes the higher order magnetic anisotropies [115], for which one anticipates the modification in spin waves although not yet shown. This strain-driven layered AFM to FM transition further enables the strain-programable magnetic tunneling junction, where the CrSBr serves as the tunneling barrier and its high and low tunneling resistance states are switched via the tensile strain [164].

***Hydrostatic pressure.*** While the uniaxial strain makes a significant impact on the interlayer exchange coupling, hydrostatic pressure aims to affect the intralayer exchange coupling. A compression of the lattice under hydrostatic pressure is shown to reduce the magnetic critical temperature at a rate of -12.6 K/GPa because of the suppression of intralayer FM coupling, while the intralayer FM and interlayer AFM orders are maintained within the achievable pressure range (**Figure 5b**) [146]. The overall impact of hydrostatic pressure is less prominent as compared to the uniaxial strain, for the case of CrSBr.

***Twisted superlattice.*** The twist between two atomic layers produces a superlattice out of the interference between the two angular mismatched atomic lattices. It has been introduced to tune the charge degree of freedom in twisted graphene and twist transition metal dichalcogenide structures [165–168] and more recently, applied to modify the spin degree of freedom in twisted 2D magnets [169–172], all of which have been limited to honeycomb lattice. Twisted CrSBr bilayers differ from them, because of their orthorhombic atomic lattice. It has already been shown in 90° twisted CrSBr bilayers that metamagnetic phases develop, showing hysteric behaviors and multistep magnetization switching under the magnetic field – which are features absent in natural bilayer CrSBr (**Figure 5c**) [150]. Moreover, it has been recently demonstrated that twisted CrSBr features a unique all-AFM tunneling junction with a high tunneling magnetoresistance ratio [151].

***Electric field effect doping.*** The 2D nature of CrSBr flakes naturally allows for carrier doping through the electric field effect. Indeed, such an electrostatic control has been practiced in few-layer CrSBr to modulate their excitonic and magnetic properties (**Figure 5d**) [147]. Because of charge carrier doping, charged exciton complex forms in bilayer and trilayer CrSBr, and it exhibits strong coupling with the magnetic order just like the neutral exciton does. Furthermore, in the field-effect device geometry, metamagnetic phases have been predicted and observed in few-layer CrSBr [147–149].

***Chemical intercalations.*** Chemical intercalation with alkali elements can achieve a higher level of electron doping than the popular field effect electric gating can do. In the case of CrSBr, Lithium intercalation has been shown to greatly enhance the magnetic onset temperature while maintaining the triaxial magnetic anisotropy (**Figure 5e**). In addition, a quasi-1D charge density wave emerges upon Lithium intercalation and persists above the magnetic transition temperature [153]. A more recent work by the Lv and Zhao group further shows that tetrabutylammonium (TBA$^+$) interaction into CrSBr can significantly modulate the magnetic and transport properties (**Figure 5f**) [173].

***Electric field.*** The magnetoelectric effect of the layered AFM state in CrSBr allows for the electric field control over the magnetism. Taking bilayer CrSBr, the layered AFM order breaks both spatial-inversion (P) and time-reversal (T) symmetries but preserves their product (PT) symmetry, allowing for the direct coupling between the P-odd electric field and this PT-symmetric AFM order. While this has not been demonstrated in bilayer CrSBr, it has been shown in bilayer CrI$_3$ that also hosts the PT-symmetric layered AFM order [174,175].

***Magnetic field.*** The magnetic field can be used to compete with the interlayer AFM coupling and the triaxial anisotropy in CrSBr, as well as modifying the geometric frustration and suppressing the spin fluctuations in CrOCl. In the case of natural CrSBr, it drives the layered AFM to FM spin flip transition with the magnetic field along the *b*-axis to overcome the interlayer AFM coupling and causes the spin canting transitions into the FM state with the field along the *a*- and *c*-axes to counteract the magnetic anisotropies (**Figure 5g**) [80,176,177]. Under the applied magnetic field, metamagnetic phase transitions can also be induced in CrOCl (**Figure 5h**).

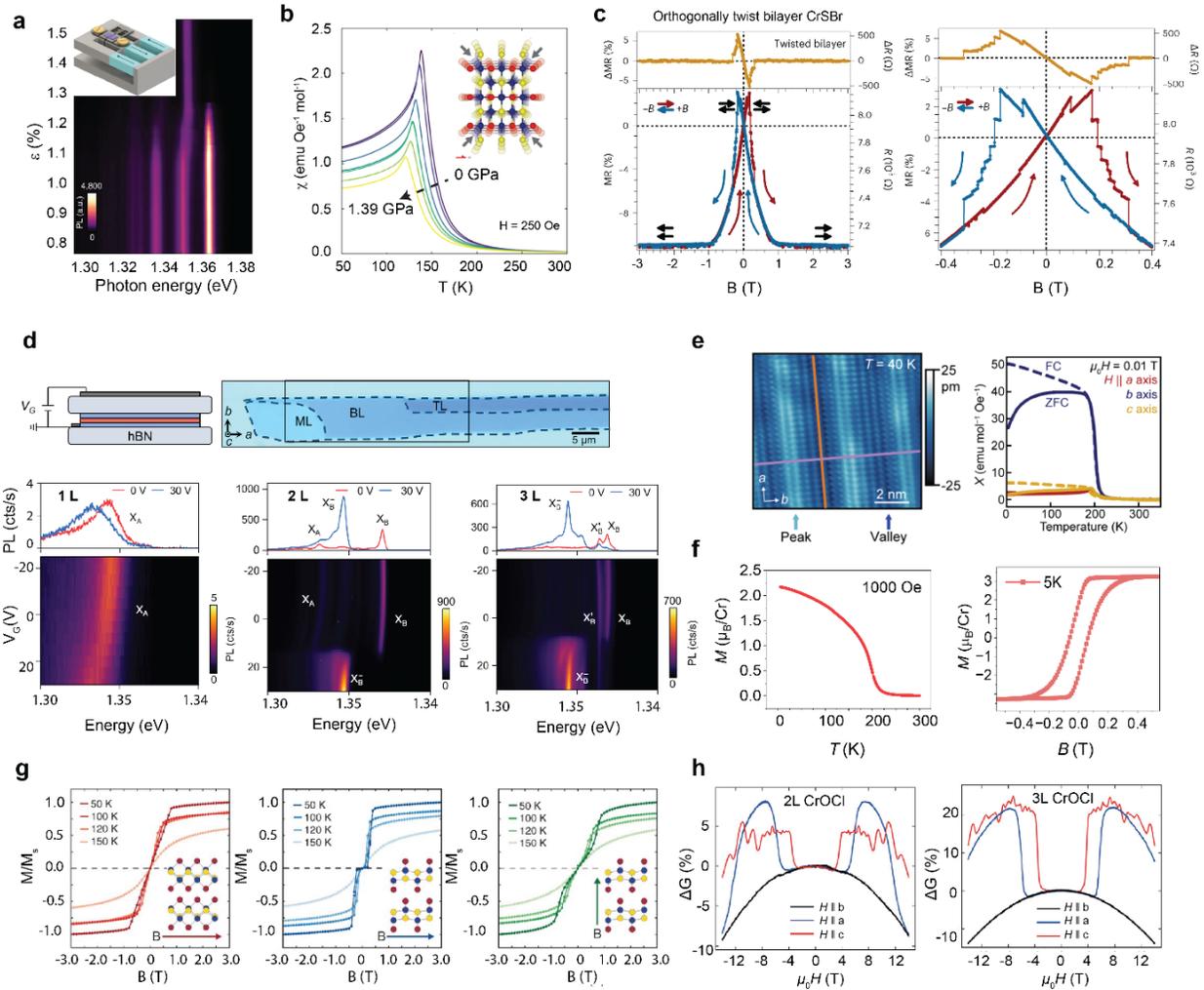

**Figure 5 | Designed controls over MXY vdW magnets. a.** Static strain control over the magnetism and excitons in CrSBr thin layers [115]. **b.** Hydrostatic pressure control of the magnetic properties in bulk CrSBr [146]. **c.** Twisting engineering the magnetic properties in twisted CrSBr homostructures [150]. **d.** Electrostatic doping effect in modifying the magnetism and excitons in CrSBr flakes [147]. **e.** Li intercalation for introducing charge density wave and increasing the magnetic onset temperature in Li-intercalated CrSBr bulk [153]. **f.** Organic molecule intercalation for altering the interlayer magnetic coupling from AFM to FM and enhancing the magnetic onset temperature [173]. **g.** Magnetic field-induced magnetic phase transitions in CrSBr bulk [80]. **h.** Metamagnetic phase transitions induced by applied magnetic fields in 2L and 3L CrOCl [113]. The figures in (**a-h**) are adapted from Refs. [115,146,150,147,153,173,80,113].

## Challenges and outlook of MXY 2D vdW magnetism

The unique lattice characters of MXY vdW magnets, *i.e.*, distorted low-symmetry crystal field and orthorhombic anisotropy atomic lattice, bestow intriguing electronic, magnetic, and optical properties that can be exploited for future spintronic, magneto-optical, or opto-magnetic devices. Yet, these prospects of MXY vdW magnets heavily rely on the crystalline quality, the comprehensive understanding, and the integration flexibility of MXY. In this section, we comment on existing challenges in these three aspects.

***Crystalline quality.*** In general, the crystalline quality of 2D magnets is poorer than that of graphene and transition metal dichalcogenide. For MXY, there are several types of defects of concern. First, the halogen deficiency often happens, for example, the Br vacancies in CrSBr [156,159]. These defects may act as trapping sites for charge carriers, and therefore, make the electric doping inefficient. They also interrupt the magnetic exchange pathways and impact the magnetic properties. Such halogen defects could be even more severe in the oxybromide and oxyiodide due to the large size and electronegativity difference between oxygen and Br/I. Second, the halogen and chalcogen inter-site mixing could happen, especially for those oxyhalide pairs with rather close anion size and electronegativity such as Se-Br, S-Cl and Te-I. In fact, site mixing has been observed in $Cr_3Se_2Br_5$ [178] and $Bi_4Cl_2S_5$ [179]. which might be the root cause preventing the successful single crystal synthesis of CrSeBr and CrSCl. Finally, the mixed phases of different chemical ratios can happen, for example, for the case of VOCl with the possibility of $VOCl_2$, $VO_2Cl$ and $VOCl_3$ [180]. It is worth noting that defects play a more profound role in magnetism, or more generally, ordering phenomena, in 2D than 3D. Therefore, novel crystal growth strategies and dedicated materials handling processes have to be developed to ensure the access of intrinsic properties of 2D MXY magnetism and to allow for efficient electrostatic device controls.

***Comprehensive probes.*** Currently, 2D magnetism is primarily probed by optical spectroscopy and spin transport measurements, mainly because the small sample volume excludes many bulk probes. For example, in 2D CrSBr, the magnetic states are mostly inferred from the exciton spectra relying on the coupling between the exciton and magnetic order [91,92,140,144,147], and sometimes analyzed through the second harmonic generation thanks to its sensitivity to point symmetries [61,82,83,177]. However, a comprehensive understanding of magnetism requires the identification of its accurate space group (including both point and translational symmetries), its complete spin wave dispersion, its real space imaging of spin textures, magnetic domains, and domain walls, *etc*. which optics and spin transport cannot provide. It calls for techniques of spin sensitivity and with momentum and/or spatial resolutions to complement optics and transport in the research of 2D magnetism including and beyond MXY. For the momentum resolution, resonant x-ray spectroscopy is gradually applied in studying 2D magnets [181]. For the spatial resolution, scanning NV microscopy [182,183], spin-resolved scanning tunneling microscopy [184], Lorentz transmission electron microscopy [185] and phase-resolved second harmonic generation [177] are introduced in imaging 2D magnets lately.

***Integration flexibility.*** It is highly desirable to harvest the intriguing physical properties of MXY by embedding them into devices. MXY is especially appealing in this regard, because first their properties can support a diverse variety of devices, including spintronics, magneto-optics, opto-magnetics, and even optoelectronics [151,186,187]. Second their vdW 2D nature allows for ultraclean interfaces and ultracompact arrangement of the device structures. And third their air stability is superb compared to many other 2D magnets. However, it is nearly impossible to scale up if relying on mechanical exfoliation to provide the 2D MXY vdW flakes. Yet, there have been no successful efforts on large-scale thin film growth even for CrSBr, the most investigated MXY family member. This possibly roots in the complication involved in handling the potentially toxic and corrosive precursors and gases during the film deposition. On the other hand, large area highly crystalline thin film growth of oxyhalides appears to be feasible with solid state metal oxides and metal halides under reduced oxygen atmosphere, as it has been demonstrated in the similar phase BiOI and LaOCl thin film growth recently [188–190].

## Acknowledgements

L. Z. and X. G. acknowledge the support from the Department of Energy Office of Basic Science under Award DE-SC0024145 (for the discussions on twisted and moiré magnetism), the National Science Foundation under Award DMR-174774, DMR-2103731, and MRSEC DMR-2309029, the Air Force Office of Scientific Research under Award FA9550-21-1-0065, the Office of Naval Research under Award N00014-21-1-2770, the Gordon and Betty Moore Foundation under Grant N031710, and the Alfred P. Sloan Foundation (for the discussions on natural 2D magnetism). B. L. and W. L. acknowledges the support from US Air Force Office of Scientific Research Grant No. FA9550-19-1-0037, National Science Foundation-DMREF-2324033 and Office of Naval Research grant no. N00014-23-1-2020 and N00014-22-1-2755.

## Author Information

These authors contribute equally: Xiaoyu Guo, Wenhao Liu.

Authors and Affiliations

**Department of Physics, the University of Michigan, Ann Arbor, MI, 48109, USA**
Xiaoyu Guo, Liuyan Zhao

**Department of Physics, the University of Texas at Dallas, Richardson, TX 75080, USA**
Wenhao Liu, Bing Lv

Corresponding Authors

Correspondence to Liuyan Zhao (lyzhao@umich.edu), or Bing Lv (blv@utdallas.edu)

## Competing Interests

The authors declare no competing interests.